# Implementation of Noise-Shaped Signaling System through Software-Defined Radio

Junsung Choi[1], Dongryul Park[1], Suil Kim[2], and Seungyoung Ahn[1,*]

[1] The CCS Graduate School of Green Transportation, Korea Advanced Institute of Science and Technology (KAIST), Daejeon 34141, South Korea; choijs89@kaist.ac.kr (J.C.); dongryulpark@kaist.ac.kr (D.P.)
[2] Agency for Defense Development (ADD), Daejeon 34186, Korea; sikim777@add.re.kr (S.K.)
* Correspondence: sahn@kaist.ac.kr

**Abstract:** As developments of electromagnetic weapons, Electronic Warfare (EW) has been rising as the future form of war. Especially in wireless communications, the high security defense systems, such as Low Probability of Detection (LPD), Low Probability of Interception (LPI), or Low Probability of Exploitation (LPE) communication algorithms, are studied to prevent the military force loss. One of the LPD, LPI, and LPE communication algorithm, physical-layer security, has been discussed and studied. We propose a noise signaling system, a type of physical-layer security, which modifies conventionally modulated I/Q data into a noise-like shape. For presenting the possibility of realistic implementation, we use Software-Defined Radio (SDR). Since there are certain limitations of hardware, we present the limitations, requirements, and preferences of practical implementation of noise signaling system, and the proposed system is ring-shaped signaling. We present the ring-shaped signaling system algorithm, SDR implementation methodology, and performance evaluations of the system by the metrics of Bit Error Rate (BER) and Probability of Modulation Identification (PMI), which we obtain by Convolutional Neural Network (CNN) algorithm. We conclude that the ring-shaped signaling system can perform a high LPI/LPE communication function due to the eavesdropper cannot obtain the correct used modulation scheme information, and the performance can vary by the configurations of the I/Q data modifying factors.

## 1. Introduction

As developments of electromagnetic weapons based on such technologies, electronic equipment and modern technologies, Electronic Warfare (EW) has been rising as the future form of wars [1, 2]. In EW, the attackers' purposes are capturing military secrets, malfunctioning core electronic equipment such as radios and radars, or spying satellites through Electromagnetic Pulse (EMP), and High Power Microwave (HPM). Leakage of military secrets can directly cause a significant loss of military forces. To defend message leakage, more complex and robust defense systems are required such as Low Probability of Detection (LPD), Low Probability of Interception (LPI), or Low Probability of Exploitation (LPE) communication algorithms [3-6]. LPD defines as the eavesdropper that cannot detect a signal at all. LPI defines as the eavesdropper detecting the presence of a signal, but does not identify the characteristics of the signal. LPE defines as the eavesdropper detecting the presence and identifies the characteristics of a signal, but cannot exploit the message. These solutions are processed at a physical-layer level. The physical-layer processing is studied to reduce the complexity by hiding the signal with dummy noises [7] or a dirty constellation algorithm; however, there is a limitation as not being able to shape as a complete noise.

Various security methods in the communication field have been proposed. The cryptography methods such as symmetrical and asymmetrical key encryption are used in wireless communication systems [8, 9]. However, these methods may generate large delays due to complex procedures with using additional bits for security. Therefore, physical-layer security processes are discussed and studied to overcome the disadvantages [10-14]. However, still, a complex and long sequence, which can occur computational overhead, is required to use for strong security performances.

In this paper, we propose a flexible and adaptable covert communication algorithm through a noise shaping process in physical-layer to overcome the implementation complexity and hardware dependency. The proposed algorithm focus to modify In-phase and Quadrature (I/Q) form of data. The proposed algorithm is composed of shared noise-shaping modifying factors, which are shared between the designated transmitter and receiver. The transmitter reshapes to a noise shape by the shared noise shaping modifying factors to original I/Q data after the conventional modulation process. The receiver or an attacker, a third party, will detect the signal as a noisy signal. Since the receiver knows the shared factors, the receiver can recover the correct I/Q data while an attacker cannot.

Most of the previous works are finished their studies at the simulation level, not considering the real implementation. We present a practical implementation of noise signaling through SDR radios which can flexibly function by installed user-customizable software. With the implementation process, we present the realistic possibility to implement the noise signaling system. As mentioned in [15, 16], the ultimate physical-layer security may be achievable when the noise shaping is much similar to Additive White Gaussian Noise (AWGN). However, due to the limitations of radio hardware, the perfect AWGN shape is not possible; certain boundaries to optimize for achieving both good security and recovered performance. Therefore, we decide to shape I/Q constellation data to ring shape rather than full AWGN.

With implemented ring-shaped signaling system SDR radios, we evaluate with performance metrics as Bit Error Rate (BER) and Probability of Modulation Identification (PMI). BER is calculated the differences between transmitted data bits in the transmitter and recovered data bits in the receiver. PMI is the probability of identifying a certain modulation scheme at the attacker's viewpoint. We obtain PMI from Convolutional Neural Network (CNN) algorithm, which is trained by BPSK, QPSK, 8~64PSK, 8~64QAM with 0~30dB SNR AWGN channel I/Q constellation data. From BER, we evaluate how well noise signaling recovery can be. From PMI, we evaluate how well ring-shaped signaling system can hide a modulation scheme from an eavesdropper.

This paper is organized as follows. In Section 2, we provide a literature review of previous studies about covert communication techniques. The presentations of limitations and preferences when implementing noise-shaped signaling process to real radios in Section 3. The proposed ring-shaped signaling system algorithm is presented in Section 4. Necessarily required functions for the implementation through SDR are presented in Section 5. Performance evaluations with BER and PMI are shown in Section 6. Lastly, the conclusion with a discussion of future works is drawn in Section 7.

## 2. Related Works

The importance of data defense from attackers is more emphasized and studied recently. The authors of [3] improve the security by adding a variable cyclic prefix (CP) and frequency domain jitter. The authors of [17] propose a random user selection method. The authors of [18] propose precoding-aided spatial modulation to improve covert performance.

The spread spectrum technology is commonly used to improve covertness [1], [19]. The spread spectrum technologies including Direct Sequence Spread Spectrum (DSSS), Frequency Hoping Spread Spectrum (FHSS), and combinations of DSSS and FHSS are usually used with CDMA and Orthogonal Frequency-Division Multiplexing (OFDM) technologies for multiple access. Moreover, various studies have been presented to enhance the advantages of the spread spectrum with CDMA [5, 6, 20]. However, these algorithms have a limitation that decoding the original message is possible when the signal is detected.

To overcome the problem, physical layer security methods have been studied with wireless communication technology. The authors of [4] present a method to improve LPI performance in physical-layer. The dirty constellations algorithm is introduced as a modulation method that adds an additional covert Quadrature Phase-Shift Keying (QPSK) signal to the original modulated signal with a low information transmission rate such as Binary Phase-Shift Keying (BPSK) and QPSK. This method mimics wireless communication signals with high noise and hides the signal at a high Error Vector Magnitude (EVM) point. Therefore, the dirty constellation of the transmission signal has a high covert performance. However, the generated constellation by the presented algorithm does not deviate from the form of a conventional QPSK; the attackers are likely to intercept and possible to decode the original message. A multiple modulation schemes broadcasting method is studied to enhance security performance compared to a single modulation scheme system [21]. A continuously varying modulation scheme is proposed [22, 23]. A method that shaping the symbol similar to AWGN is also proposed [24]. However, the proposed methods have limitations that the modulation scheme may easily be guessed and require dummy data.

In this paper, we propose a ring-shaped signaling system algorithm that cannot demodulate the message even if the transmitted signal is detected and intercepts. The algorithm is mainly focused on LPI/LPE communication function. In this algorithm, additional shared modifying factors are used for shaping and recovering the signal. This method has the advantages of being easy to implement, adaptable for any communication system using I/Q data, and effectively decreasing the attackers' probability to decode.

## 3. Noise-shaped Signaling Definition, Limitation, and Preferences

The purpose of using noise-shaped signaling is to achieve signal transmission security. 3 categories of the noise-shaped signaling function are discussed: 1) eavesdropper is not able to detect the transmitting RF signal, LPD, 2) able detect the presence of transmitting RF signal, but not able to interpret the signal characteristics, LPI or 3) even interpret the signal, not able to exploit due to unable to decode the correct message, LPE. The ultimate goal of noise-shaped signaling is transforming the I/Q signal distribution into almost equal to Gaussian distribution. As the signal is transformed into almost equal to Gaussian noise, the RF signal may achieve the highest security, achieving LPD, LPI, and LPE functionalities. However, when trying to implement the signaling methods to the real radio, the receiver may face the most difficult condition to decode the RF signal due to several hardware limitations.

When adapting the Gaussian distributed noise-shaped signaling through SDR, there are several considerable issues such as signal power control, signal synchronization, additional noise mixing process, etc.

For signal power control issues, the usual radios have the maximum transmission power limitation for several reasons as same as SDR. Also, SDR receivers control their own gain to protect from the over-powered received signal or increase the performance when the transmission power is too low. Therefore, the noise-shaped signaling process also requires following the maximum power limitation rule. The reshaped signals' power should not exceed the maximum power of the original signal, requiring the highest bound. Also, the lower power matters to the radio performance as well. For correctly decoding the messages, a certain value of decision distances between I/Q data is required. As lowering the received power, the decision distances decrease as well and can defect the performance of the radio. Therefore, considering the possibility of critical additional noise appearance between the communication channel, the noise-shaped signaling should have the lowest bound.

The basic concept of noise-shaped signaling is modifying the original signal into a noise-like form. The process may be done by addition, subtraction, multiplication, or division of the original signal and modifying parameters. When the receiver recovers the original signal from the received signal which is affected by the noise-shaped signaling and channel, the receiver should require almost perfect synchronization in both frequency and time domain for neglecting the noise-shaped signaling effect and only facing the channel effect. Therefore, the communication system should have a function for perfect synchronization.

When the additional noise mixing process is proceeding, one of the following calculations is used: addition, subtraction, multiplication, or division. Also, the mixing process may directly apply to data bit, I and Q data at the same time, or independently. In SDR, the transmitter and receiver gain can be set up differently, which can change I/Q data magnitude. Therefore, the addition or subtraction process may not feasible when implementing through SDR. Also, synchronization functions are using inputs as I/Q data form. So that for achieving the purpose of securing modulation scheme information, applying the noise mixing process at I/Q data, after modulation function at the transmitter and before demodulation function at the receiver, is adapted.

With the observed definitions, limitations, and preferences, we propose ring-shaped signaling system algorithm as the suitable noise signaling algorithm for implementing through SDR. The detailed descriptions of the algorithm, implementation processes, and performance analyses are presented in the following sections.

## 4. Ring-Shaped Signaling System Algorithm

The proposed ring-shaped signaling system block diagram is shown in Figure 1. After the conventional modulation process of the communication system, a noise-shaping operation is added to modify a signal having the characteristics of noise shape. The envelope that using for noise-shaping operation is composed of phase and magnitude, and it is possible to control the noise level by different variables. The phase envelope variable rotates the constellation. For example in PSK signal, the result of the rotation process is modifying the PSK constellation to a different order of PSK signal which makes it difficult for attackers to interpret the data and look like a noisier signal when the rotation order increases. The magnitude envelope displaces the original I/Q constellation data points to close to the origin point (I/Q = (0,0)). The transformed signal by magnitude envelope looks more like white noise, and the magnitude modification process lowers the power of the signal which is difficult to intercept.

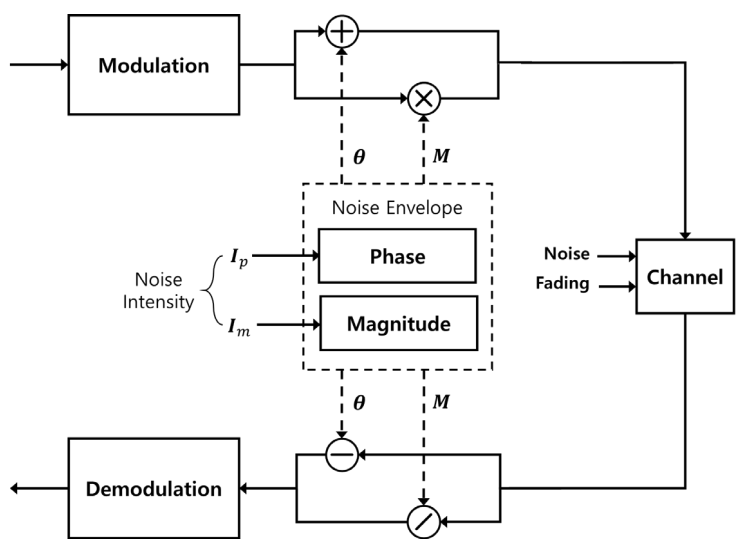

**Figure 1.** Ring-shaped signaling system block diagram.

The expected modified constellation through the proposed ring-shaped signaling system is shown in Figure 2. With the proposed system, the data points around the origin (I/Q = (0,0)) are displaced with a certain low boundary to form a ring. By the ring-shaped signal, the constellation seems faded or white noise with a hole at the center. Therefore, the eavesdropper is difficult to interpret the data because of unable to notice the used modulation scheme. In addition, since the modified signal is similar to noise, the average transmission power of the signal is relatively lower than the original signal, the lower transmission power also able to decrease the probability to intercept the signal.

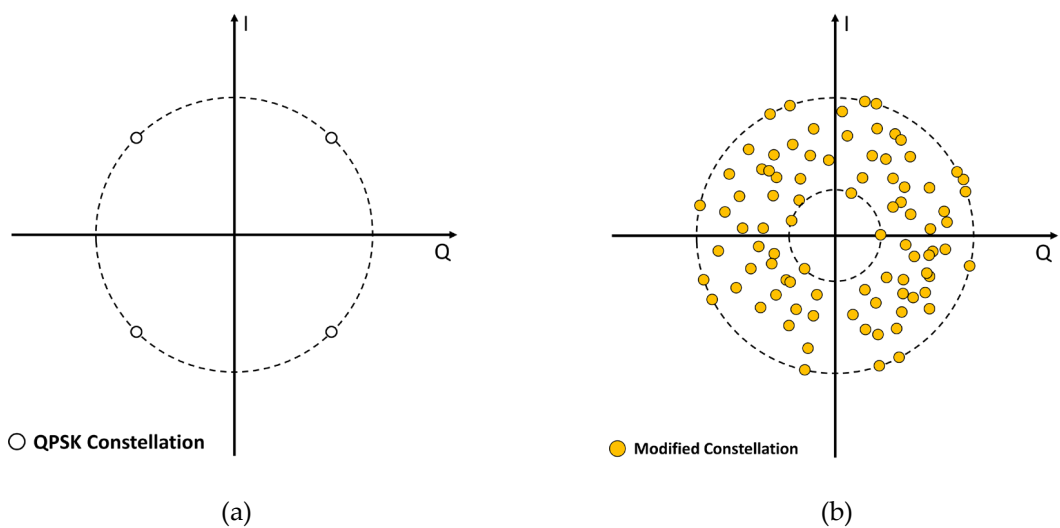

(a) (b)

**Figure 2.** I/Q constellation of: (**a**) before signaling process; (**b**) after ring-shaped signaling process.

The ring-shaped signaling system converts a conventionally modulated signal as shown in equation (1) into a modified signal by a magnitude modifying factor and a phase modifying factor as shown in equation (2). The magnitude modifying factor and the phase modifying factor can adjust the noise level through independent variables, and the performance of covertness can be changed according to the factors.

$$S = \chi(\cos\varphi + j\sin\varphi) \tag{1}$$
$\chi$: magnitude of signal

$$S_n = \chi M(\cos(\varphi + \theta) + j\sin(\varphi + \theta)) \tag{2}$$
M: magnitude modifying factor
$\theta$: phase modifying factor

The phase modifying factor is generated through the process as shown in Figure 3. After generating random number bits, $n$, the random bits are converted to a decimal number according. The converted decimal is divided by the value of the intensity level, $I_p$, multiplier of 2. Through this process, the range of $(0, 2\pi)$ is divided by a certain resolution according to the intensity level. The modified signal is rotated as shown in Figure x2 (b).

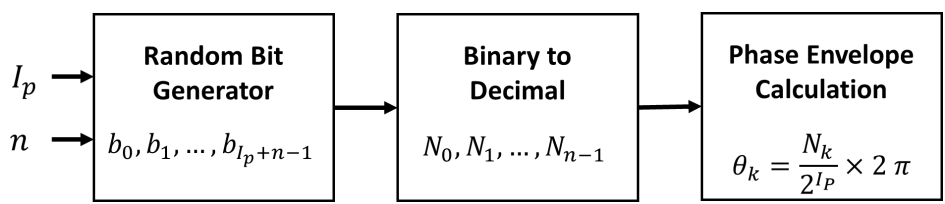

**Figure 3.** Phase modifying factor generation process.

The magnitude modifying factor is applied to the original signal by multiplying a uniformly distributed random number that has an equal length as the transmitted data. As shown in Figure 4, when generating the magnitude modifying factor, the minimum value of the random number is adjusted according to the magnitude intensity level, *I*<sub>*m*</sub>, to control the lower boundary that influences the signal magnitude.

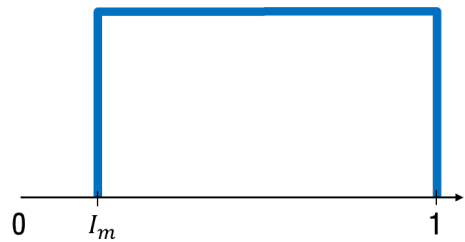

**Figure 4.** Distribution of magnitude modifying factor.

The modified signal by the phase modifying factor and magnitude modifying factor is shown in Figure 5 (a) and (b), respectively. The signal modified by the phase modifying factor contains data points which evenly distributed in certain degrees of angles according to the resolution of the intensity level and shape like a circle as shown in Figure 5(a). The signal modified by the magnitude modifying factor has data points that are distributed with different magnitude values; they are displaced toward inside the circle as shown in Figure 5(b).

Ring-shaped signaling system that simultaneously applied with phase modifying factor and magnitude modifying factor has lower average signal energy than the original signal due to the reduced average magnitude as shown in equation (3). The reduction of signal energy is possible to increase covertness performance, but the performance of the recovered communication system degraded because of more sensitive to the channel noise; reduced magnitude is directly related to the reduced decision-making distance as shown in equation (4). Therefore, the noise signaling system must be driven by adjusting the suitable parameters for the best performance in the recovered communication system and covertness at the same time.

$$P_{ring} = \frac{1}{T} \int_0^T \left| \frac{I_m + 1}{2} A e^{j\omega t} \right|^2 dt = \left( A \frac{I_m + 1}{2} \right)^2 \qquad (3)$$

$$BER_{ring} = \frac{\sum_1^{symbol\ length} 2Q\left( \sqrt{\frac{\left(\frac{I_m + 1}{2}\right)^2 E_0}{N_0}} \right)}{symbol\ length} \qquad (4)$$

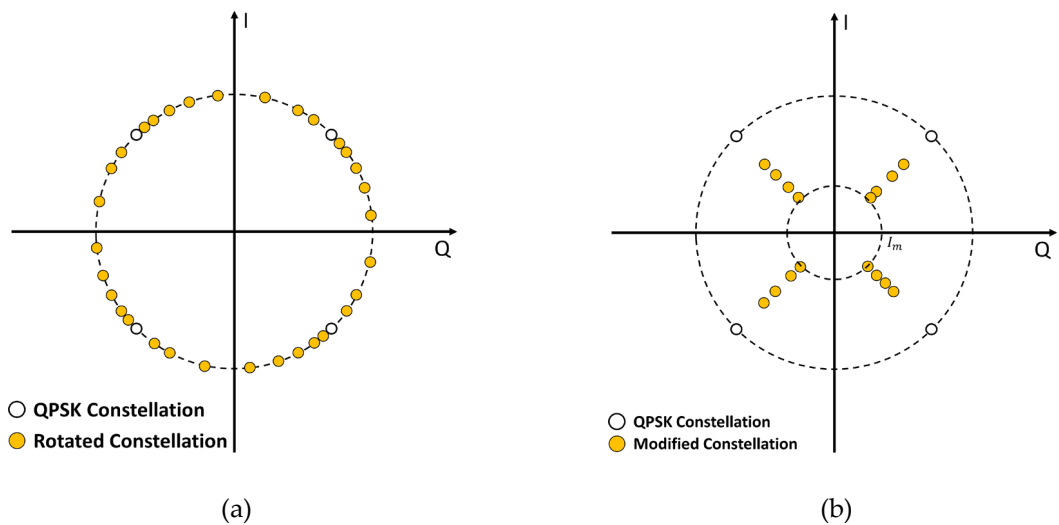

(a) (b)

**Figure 5.** Ring-shaped signaling processed I/Q constellation of: (a) only affected by phase modifying factor; (b) only affected by magnitude modifying factor.

## 5. Required Functions for SDR Implementation

We try to implement the ring-shaped signaling system through SDR. Due to several configuration complexities, we focused only PSK transmitting and receiving communication system; evaluating whether effectively covert QPSK system without performance degradation. We used Ettus USRP B210 for SDR and Matlab for software.

When the ring-shaped signaling system is implemented through SDR, there are several important functions. Because the conventional QPSK signal shape changes to a signal with a different modulation order due to the phase modifying factor, any frequency-related functions are required among the transceiver.

For the data recovery, accurately extracting the envelope is necessary. To accurately eliminate the envelope values to each data, the accurate synchronization between the transmitted and received data is needed for aligning the data in the correct order. For accurate synchronization, pre-processes with RF signal by using certain information such as modulation, a sample rate of a transmitted signal, or data header are required.

Within the RF signal pre-processing, we consider the required functions that are affected according to modulation scheme types are the carrier synchronizer function and the frequency offset estimator function. The carrier synchronizer function uses the Phase Locked Loop (PLL) method. Since the PLL method depends on a different step of the phase error value, the transceiver must share the phase information of the same MPSK modulation order for correctly estimating the phase error step. In the case of using the ring-shaped signaling system, the received RF signal is changed to a certain order of MPSK which is changed by the phase modifying factor. Since the order is flexibly changed, the PLL setting of the receiver must match with the order of MPSK which changes flexibly. Similarly, the frequency offset estimator function must also be adjusted to the changed PSK order for the same reason. Carrier synchronizer function and frequency offset estimator function are shown in Figures 6(a) and (b) below.

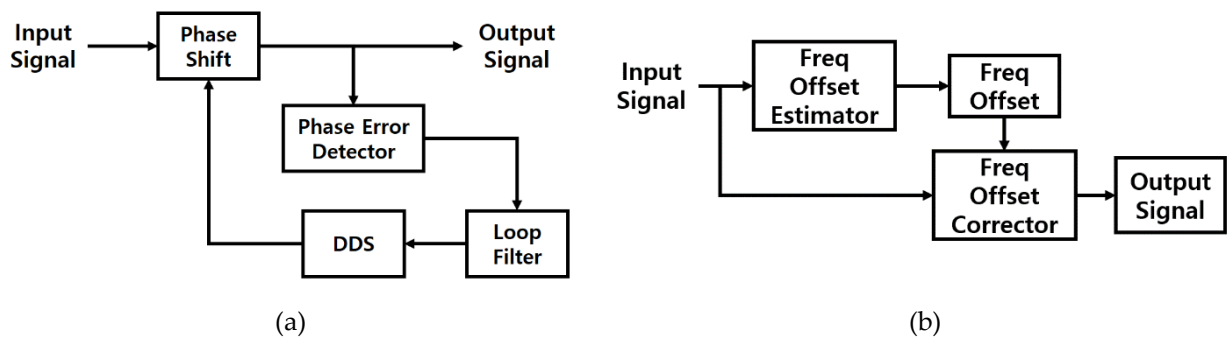

(a)     (b)

**Figure 6.** Block diagram of: (a) Carrier synchronizer function; (b) frequency offset estimator function.

In the conventional communication system, the symbol length depends on the type of modulation scheme and is calculated as shown in Figure 7, dividing the bit length by $\log_2$ of the modulation scheme. This calculation process is also applied to both header, data, and tail bits. The converted symbol uses for forming RF signals in transceivers. In the conventional communication system, the symbol form of header uses the input for the carrier synchronizer and frequency offset estimator functions.

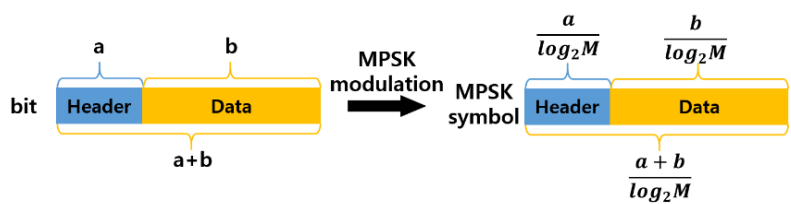

**Figure 7.** Conventional bits to the symbols transformation process.

In general, if the length of the header changes, it is difficult to find the exact starting point of the data so that accurate alignment is not possible. Therefore, we propose a symbol generation process that generates symbol form of data and header independently. The process first uses conventional symbol data transformation process only for the data bit part, generating predefined header symbol, and then merging the two-part symbol data to generate a full form of symbol data. For accurate synchronization, the symbol form of the header is chosen for 180° apart symbols. Since the separately generated symbol form header can have the same length no matter what the phase modifying factor is used. With the same shared header, the accurate data alignment for the synchronization can be obtained even if the PSK modulation order is changed over a flexibly changing phase of the ring-shaped signal.

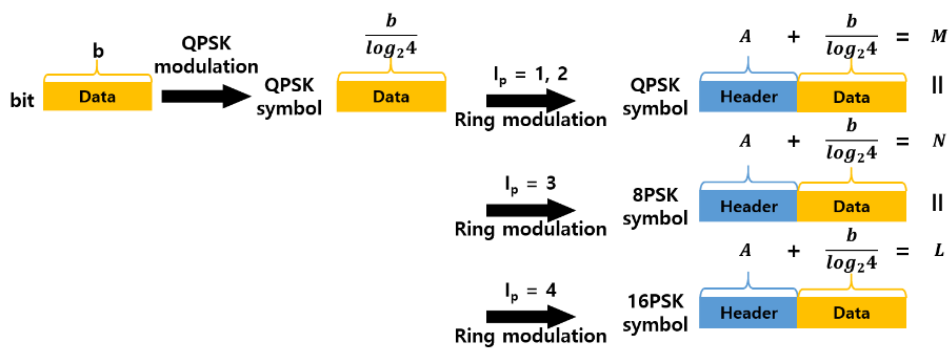

**Figure 8.** Proposed header and data separated bits to the symbols transformation process.

## 6. Results

### 6.1. SDR Performance Evaluations

The SDR transceiver experiment environmental setup is shown in Figure 9. It consists of a transmitter, a laptop that controls the SDR transmitter, a receiver, and a laptop that controls the SDR receiver. The laptop uses Intel Core i7-8750H processer and 16GB RAM. The used SDRs are Ettus USRP B210. In addition, for minimizing the external noise effect, we connect the RF port through a cable during the performance evaluation experiments. The detailed SDR configuration setups are shown in Table 1.

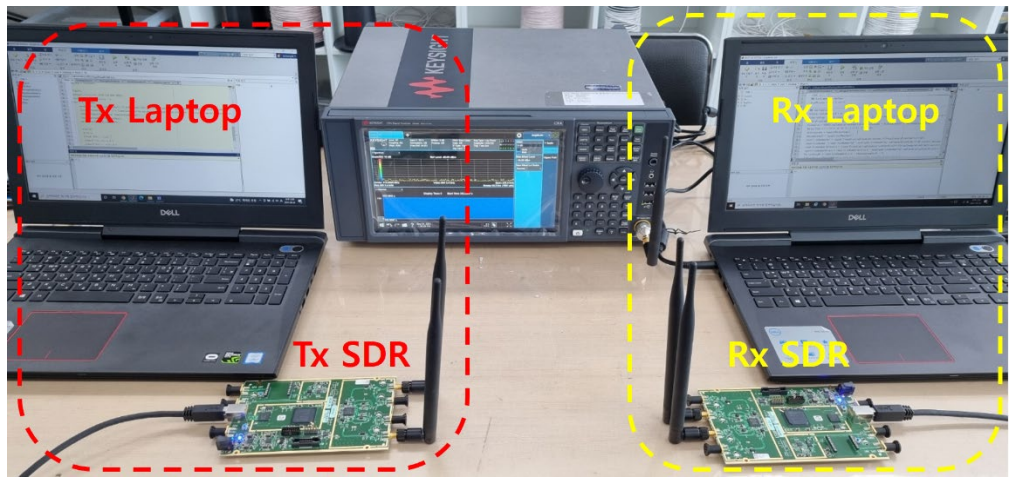

**Figure 9.** SDR performance evaluation experiment setup.

**Table 1.** SDR configuration parameters.

| Parameter Name | Parameter Values |
|---|---|
| Center frequency | 915MHz |
| Bandwidth | 1MHz |
| Data bit | 10,000 |
| Clock rate | 20MHz |
| Header symbol length | 12 symbols |
| Message | hello world ### |

We conducted an experiment to compare the theoretical BER and SDR performance for QPSK, 8PSK, 16PSK, which representative up to 4 different levels of phase modifying factor. As shown in Figure 10, we confirm that the trends for theoretical and SDR performance are somewhat similar to each other. However, as the modulation order increased, we observe that the differences between SDR performance and the theory are increased. Also, at the low SNR values, approximately less than 3dB, SDR performances have more error than the theory. However, at the high SNR, we solidly expect that SDR can perform as similar as the theoretical performance.

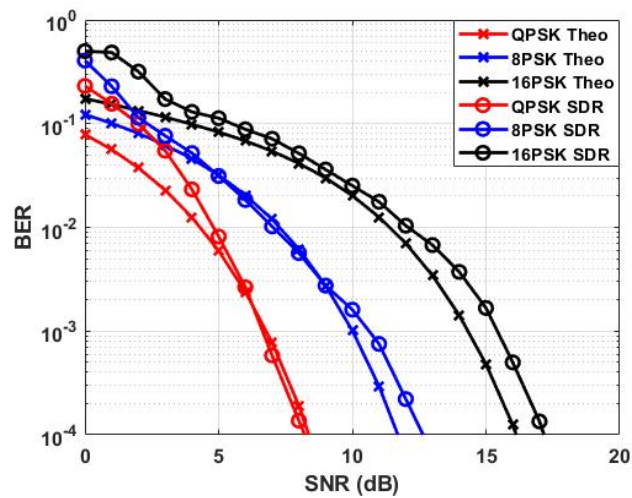

**Figure 10.** BER comparison between theoretical values and SDR performances.

Even though there are some differences observed for the BER performance, the ring-shaped signaling system has a higher priority on I/Q constellation shape rather than recovered communication performance. As shown in Figure 11, we confirm that the I/Q constellation forms of QPSK, 8PSK, and 16PSK are clearly distinguishable. From the observation, we can define that the SDR system's recovering data function is not perfectly working though I/Q constellation shaping is properly working. As we observed in the BER performance evaluation, the increased errors between the theoretical value and SDR performances according to the modulation order, the I/Q constellation data points' concentrations are more scattered as increased modulation order. With the confirmed properly working I/Q constellation up to 16PSK, the ring-shaped signaling system implementation through SDR can function up to $I_P = 4$.

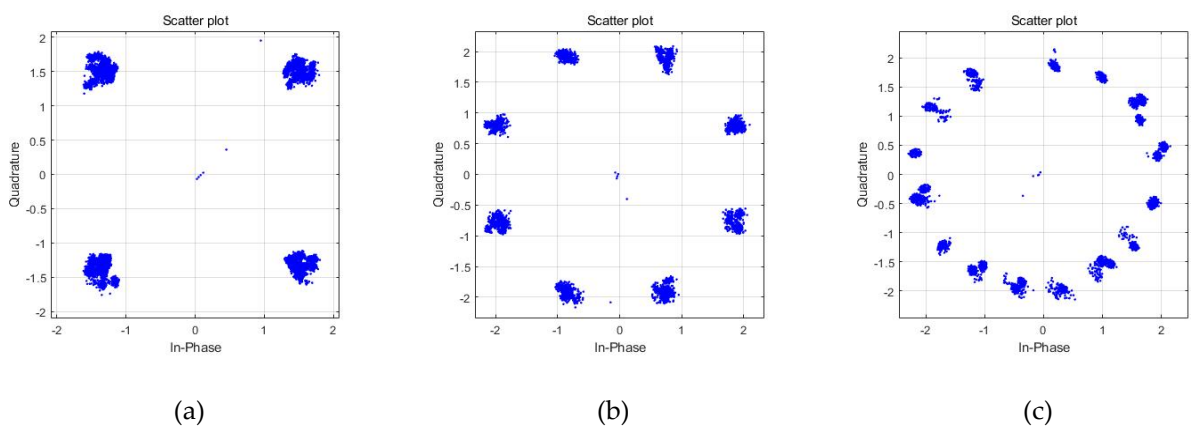

(a) (b) (c)

**Figure 11.** I/Q constellation shapes of: (a) QPSK; (b) 8PSK; (c) 16PSK.

For evaluating the ring-shaped signaling system functioning property, we conduct transceiver performance evaluation experiments with varying the phase and magnitude modifying factors. As shown in Figure 12(a), we observe that the performance degrades as the magnitude modifying factor decreases. As shown in Figure 12(b), we observe that the performance can be recovered even though the different phase modifying factors are applied. As shown in Figure 12(b), in the case of $I_P = 4$, 16PSK, the full recovery is not done due to the functional error existence for 16PSK modulation scheme performance in SDR without noise signaling. Therefore, we assume that if the plain SDR transceiver is properly functioning, the perfect performance recovery with the different phase modifying factors may possibly achieve.

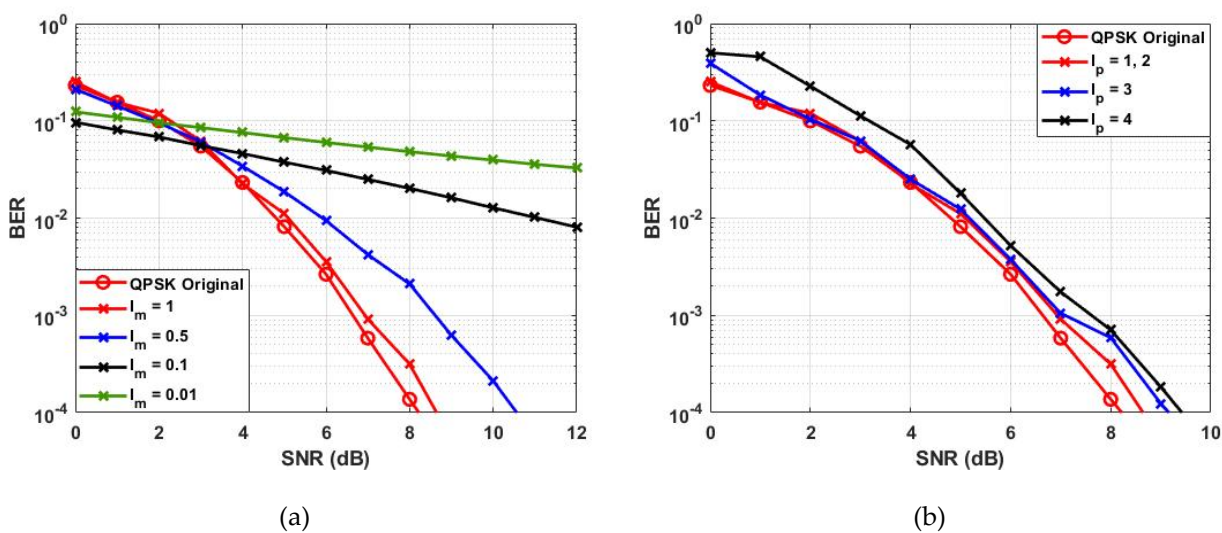

(a) (b)

**Figure 12.** Performance evaluations with varying: (a) magnitude modifying factors ($I_m$); (b) phase modifying factors ($I_P$).

*6.2. PMI Performance Evaluations*

To define the covertness performance, we conduct the modulation identification evaluations with the CNN based modulation identification algorithm. The purpose of the evaluation is to check whether a third party could identify what modulation scheme is used. The CNN algorithm is constructed and trained with BPSK, QPSK, 8PSK, 16PSK, 32PSK, 64PSK, 8QAM, 16QAM, 32QAM, and 64QAM modulation scheme data with 0~30dB SNR channel with using Matlab. With the trained CNN algorithm, we can obtain PMI as a performance metric. PMI is an indicator showing the probability of which modulation scheme is recognized; a higher value has a higher probability to be recognized.

From the modulation identification accuracy experiment, we evaluate the prediction accuracy of CNN algorithm for trained modulations. We generate 20,000 signal data per each modulation scheme. Within 20,000 signal sets, 80% of them are used for training, 10% for validation, and 10% are used for testing. As shown in Figure 13, in the case of QAMs, the probabilities are at least 80% even for the higher modulation orders. But in the case of PSKs, the probabilities are significantly lowered for the modulation order of 32 or higher. Therefore, the modulation order of up to 16PSK for PSK and any QAM can be accurately detected. So that the PMI obtained from the CNN is trustable.

**Figure 13.** PMI accuracy table of constructed CNN algorithm.

Using the developed CNN algorithm, we evaluate which modulation is recognized at the eavesdropper point of view when the ring-shaped signaling system is applied to a conventional QPSK system; we vary the magnitude and phase modifying factors. Among the two factors of the ring-shaped signaling system, the change in magnitude and phase were fixed.

We conduct an experiment of $I_P$ = 1 with varying magnitude modifying factor. As shown in Figure 14, the CNN algorithm well detects the modulation scheme as QPSK for high magnitude modifying factor. But as the magnitude modifying factor decreases, the CNN algorithm recognizes as 64QAM, and further confusing which modulation scheme is used. From the experiment, we observe that the modulation scheme can be effectively hidden when $I_m$ is less than about 0.3. Also, we can conclude that the ring-shaped signaling system has LPI communication functionality due to the ability to hide the modulation scheme from the eavesdropper.

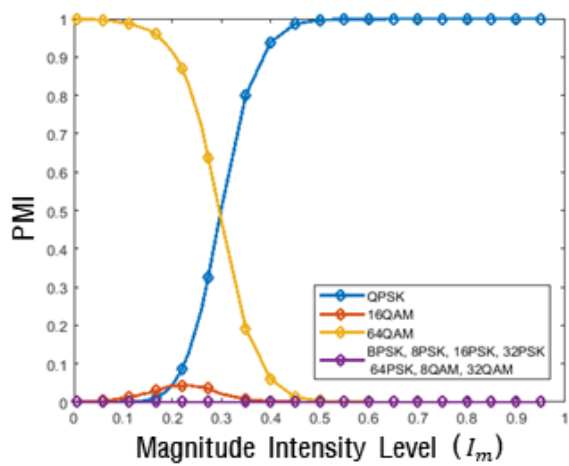

**Figure 14.** PMI evaluation with varying $I_m$ and $I_p$ = 1.

We conduct another similar experiment but modifies $I_p$ = 4. As shown in Figure 15, the CNN algorithm is not possible to infer QPSK for any scenarios. Moreover, when the magnitude modifying factor is decreased, the algorithm is recognized as QAM signals. In addition, even when the magnitude is increased, the algorithm recognizes as 16PSK rather than QPSK signal, the original modulation scheme. Therefore, we confirm that the modulation scheme cannot be inferred when the ring-shaped signaling system is applied, achieving LPE communication functionality. Furthermore, we are sure that if the phase modifying factor is large enough to scatter I/Q constellation points as white noise, the modulation scheme leakage will be impossible even for the high magnitude modifying factor, and possibly achieve both LPI/LPE communication functionality at the same time.

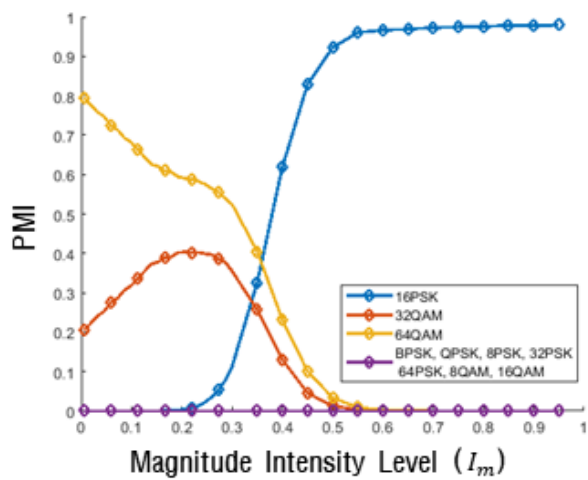

**Figure 15.** PMI evaluation with varying $I_m$ and $I_p$ = 4.

## 7. Conclusions

As EW attacks are various as developments of modern technologies, a proper defense system for a specific type of attack or a robust defense system is required. In wireless communication systems, the attackers' purposes are detecting the presence of communication, intercepting the signal characteristics, decoding the message which contains military secrets, and disabling the communication system. On the contrary, several covert communication techniques are used to defend against several attacks. Also, physical-layer defense techniques are studied for achieving LPD, LPI, and LPE communication functionalities.

From this paper, we propose the ring-shaped signaling system, a flexibly adaptable and easily implementable, which improves the security in response to LPI and LPE communication functions. The proposed ring-shaped signaling system modifies the generated I/Q signal into a ring shape by the shared magnitude and phase modifying factor. In addition, the modified performance by the proposed system is formulated from the perspective of BER and transmission power. Furthermore, we implement the noise signaling system through SDR. Also, we present the limitations, requirements, and preferences of configurations of noise signaling when implementing through SDR. The performance evaluations of noise signaling system installed SDRs are done with the performance metrics as BER and PMI.

From the evaluations, we observe that the recovered BER performance is affected by the values of magnitude modifying factors while not affected by the phase modifying factor. As reducing the magnitude modifying factor, the recovered BER performance is worse. However, PMI increases as reduced magnitude modifying factor value. At the same time, the eavesdropper cannot identify the correct modulation scheme due to the effect of the phase modifying factor, but the eavesdropper may recognize some modulation scheme information that may use to predict the original modulation scheme. From the observations, we can conclude that the combinations of using the magnitude and phase modifying factor with proper configuration value can achieve a strong LPI/LPE security performance and good recovery performance at the same time.

For future work, the following studies can advance noise signaling researches. Using channel states to extract pre-shared modifying factors can solve the limitation of our requirement, the transmitter and receiver must share the modifying factors. We only consider one-to-one communication networks; however, since the usual communication networks use multiple radios, we need to consider Media Access Control (MAC) as well as the upper layer processes. Even we present the practical implementation through SDR, it is still a basic presentation. Therefore, further implementation with commercialized products and further experiments in a real environment will be needed.

**Author Contributions:** All the authors have contributed to collecting results, performing analysis, and creating this article. All authors have read and agreed to the published version of the manuscript.

**Funding:** The authors gratefully acknowledge the support from Electronic Warfare Research Center at Gwangju Institute of Science and Technology (GIST), originally funded by Defense Acquisition Program Administration (DAPA) and Agency for Defense Development (ADD).

**Conflicts of Interest:** The authors declare no conflict of interest. The funders had no role in the design of the study; in the collection, analyses, or interpretation of data; in the writing of the manuscript, or in the decision to publish the results.